\documentclass{emulateapj} 
\usepackage{apjfonts} 

\usepackage{apjfonts} 
\usepackage{epsfig}
\usepackage{amsmath}
\usepackage{natbib}
\usepackage{verbatim}
\usepackage{rotating}
\usepackage{lscape}

\slugcomment{ApJL Accepted [1 June 2007]}

\shorttitle{AGN IN HIGH-REDSHIFT CLUSTERS} 
\shortauthors{EASTMAN ET AL.}

\newcommand{\etal}{{\rm et al.}}
\newcommand{\ergs}{erg s$^{-1}$}
\newcommand{\kms}{km s$^{-1}$}
\newcommand{\chandra}{{\it Chandra}}

\begin{document}
\title{First Measurement of a Rapid Increase in the AGN Fraction in 
High-Redshift Clusters of Galaxies}
\author{Jason Eastman\altaffilmark{1}, Paul Martini\altaffilmark{1}, 
        Gregory Sivakoff\altaffilmark{1}, Daniel D. Kelson\altaffilmark{2}, 
        John S. Mulchaey\altaffilmark{2}, and Kim-Vy Tran\altaffilmark{3,4}}

\altaffiltext{1}
{Department of Astronomy, 
The Ohio State University,
140 W.\ 18th Ave., 
Columbus, OH 43210; 
jdeast,martini,sivakoff@astronomy.ohio-state.edu}

\altaffiltext{2}
{Carnegie Observatories, 
813 Santa Barbara Street, 
Pasadena, CA 91101-1292;
kelson,mulchaey@ociw.edu}

\altaffiltext{3}
{Leiden Observatory,
Niels Bohrweg 2,
2333 CA Leiden, 
The Netherlands;
kimvy.tran@gmail.com}

\altaffiltext{4}
{Institute for Theoretical Physics, 
University of Z\"urich, 
CH-8057 Z\"urich, Switzerland} 

\begin{abstract}
We present the first measurement of the AGN fraction in high-redshift
clusters of galaxies ($z \sim 0.6$) with spectroscopy of one cluster
and archival data for three additional clusters. We identify 8 AGN in
all four of these clusters from the \chandra\ data, which are
sensitive to AGN with hard X-ray (2-10keV) luminosity $L_{X,H} >
10^{43}$ \ergs\ in host galaxies more luminous than a rest frame $M_R
< -20$ mag. This stands in sharp contrast to the one AGN with $L_{X,H}
> 10^{43}$ \ergs\ we discovered in our earlier study of eight
low-redshift clusters with $z = 0.06 \rightarrow 0.31$($\bar{z} \sim
0.2$). Three of the four high-redshift cluster datasets are sensitive
to nearly $L_{X,H} > 10^{42}$ \ergs\ and we identify seven AGN above
this luminosity limit, compared to two in eight, low-redshift
clusters. Based on membership estimates for each cluster, we determine
that the AGN fraction at $z \sim 0.6$ is $f_A(L_X>10^{42};M_R<-20) =
0.028^{+0.019}_{-0.012}$ and $f_A(L_X>10^{43};M_R<-20) =
0.020^{+0.012}_{-0.008}$. These values are approximately a factor of
20 greater than the AGN fractions in lower-redshift ($\bar{z} \sim
0.2$) clusters of galaxies and represent a substantial increase over
the factors of 1.5 and 3.3 increase, respectively, in the measured
space density evolution of the hard X-ray luminosity function over
this redshift range. Potential systematic errors would only increase
the significance of our result. The cluster AGN fraction increases
more rapidly with redshift than the field and the increase in cluster
AGN indicates the presence of an AGN Butcher-Oemler Effect.
\end{abstract}


\section{Introduction}

Measurements of AGN in low-redshift clusters of galaxies have received
substantial recent interest because of their apparent importance for
regulating the heating of the intracluster medium
\citep{mcnamara00,fabian00} and the efficiency with which sensitive
\chandra\ observations can identify the relatively rare AGN population
in clusters. In recent work \citet{martini06} showed that the X-ray
selected AGN in clusters was 5\% (for 0.5-8keV $L_X > 10^{41}$ \ergs\ in
galaxies above $M_R < -20$ mag), or five times as many as found via
emission-line selection in purely spectroscopic surveys
\citep{dressler85}.  These observations also demonstrated that the
excess X-ray point source surface density observed toward the fields
of clusters of galaxies \citep[e.g.][]{lazzati98,cappi01,sun02} were
due to X-ray emission from cluster members and not chance
associations.

The high-redshift cluster AGN population has been less well studied
due to the absence comparably-sensitive \chandra\ observations of $z>0.5$
clusters and the greater difficulty of spectroscopic
follow-up. Nevertheless, AGN are expected to be more common in
higher-redshift clusters because of the overall increase in the AGN
space density at high redshift \citep[e.g.][]{osmer04} and the
increase in the fraction of star forming galaxies in clusters at
higher redshift \citep[e.g.][]{postman01}, which indicates that
high-redshift cluster galaxies are richer in cold gas than their
lower-redshift counterparts.  Observations of several high-redshift
clusters have also found substantial AGN populations in some clusters
\citep{dressler83,johnson03,johnson06}, while surface density studies
have found evidence for an increase in the X-ray source surface
density at high redshift \citep{dowsett05}. Feedback from cluster AGN
could also play a substantial role in shutting down star formation in
cluster galaxies, such as illustrated by the significant decrease in blue
galaxies found by \citet{butcher78}.

In this {\it Letter} we present the first measurement of the cluster
AGN fraction at high redshift ($z \sim 0.58$) from analysis of four
archival \chandra\ observations, our spectroscopic observations of one
cluster (MS2053-04), and literature data for three additional clusters
(CL0016+1619 at $z = 0.5466$; CL0542-4100 at $z = 0.630$;
CL0848.6+4453 at $z = 0.57$). These observations reveal a substantial
increase in the cluster AGN fraction at high redshift, which has many
implications for the preheating of the ICM during cluster assembly,
downsizing in the AGN population as a function of environment, and the
AGN contribution to X-ray studies of high-redshift clusters for
cosmological studies.

\section{Observations}


\begin{figure*}
\plotone{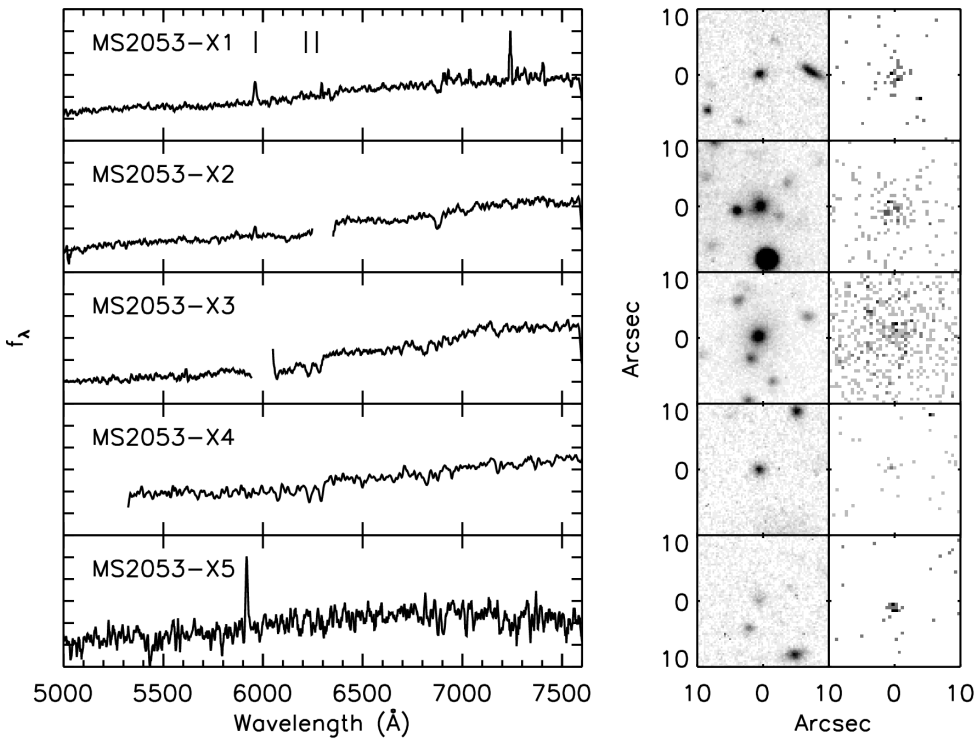}
\caption{Spectra ({\it left}), $R-$band image ({\it middle}), and
X-Ray image ({\it right}) of X-ray sources in MS2053-04.  The gaps in
the spectra correspond to the gaps between the chips of the IMACS
CCDs. Redshifted [OII] $\lambda3727$ is present in X1, X2, and X5, but
not X4, while this wavelength falls in a chip gap for X3. Ca H \& K
can be seen in X3 and X4, but not in X1 or X5 and falls on the chip
gap for X2. The location of the redshifted [OII] and Ca H \& K are
indicated in the top panel.}
\label{fig:cluster}
\end{figure*}

We have identified AGN in MS2053-04 via spectroscopic observations of
X-ray counterparts from archival \chandra\ observations. MS2053-04 has
been observed twice with \chandra\ (ObsID 551 and 1667), which are a
pair of 45ks ACIS-I observations that partially overlap one another
(both include the cluster core). The area of the first of these
datasets was imaged with deep $BVRI$ observations with the Tek5 CCD
camera at the 2.5m du Pont telescope at Las Campanas Observatory in
September 2002 (the second \chandra\ dataset was not yet
public). These images were processed and galaxies identified and
matched to the X-ray observations following the procedures described
in \citet{martini06}.

This analysis identified 82 X-ray sources with visible-wavelength
counterparts and these were used to design slit masks for the IMACS
spectrograph \citep{dressler06} on the 6.5m Walter Baade Telescope of
the Magellan Project.  A total of four, multi-tier slit masks were
designed with approximately 180 objects per mask; however, faint
sources were assigned to multiple masks and therefore a total of 617
unique objects were observed. These objects included all X-ray
counterparts brighter than $I \leq 23.5$ mag.  Additional sources were
prioritized by $I$ magnitude to obtain information about the inactive
cluster galaxy population. Color and image quality constraints were
not employed to pre-select potential cluster members due to the small
expected size of typical cluster members and the large population of
relatively blue or Butcher-Oemler galaxies in this high-redshift
cluster.

%
%
\begin{deluxetable*}{lllllllll}
\tablecolumns{9}
\setlength{\tabcolsep}{0.03in}
\tabletypesize{\scriptsize}
\tablecaption{X-ray Sources at MS2053-04 Redshift \label{tbl:ms2053} } 
\tablehead{ 
\colhead{Name} &
\colhead{Object ID} &
\colhead{I} &
\colhead{V-I} &
\colhead{log $L_{X,H}$} &
\colhead{$z$} &
\colhead{$R$ [Mpc]} & 
\colhead{$R/r_{200}$} & 
\colhead{Notes} \\
}
\startdata
MS2053-X1   & CXOU J205647.1-044407 & 21.26 (0.05) & 0.950 (0.08) & 42.80$^{+0.15}_{-0.12}$ & 0.600  & 3.57 & 2.35 & Not in 1st X-Ray Field  \\ 
MS2053-X2   & CXOU J205617.1-044155 & 18.96 (0.05) & 2.020 (0.08) & 42.81$^{+0.10}_{-0.08}$  & 0.600  & 1.67 & 1.10 &                         \\
MS2053-X3   & CXOU J205621.2-043749 & 19.20 (0.05) & 2.048 (0.08) & 42.43$^{+0.20}_{-0.16}$  & 0.584  & 0    & 0    & BCG (likely not an AGN) \\
MS2053-X4   & CXOU J205621.2-043552 & 21.35 (0.05) & 1.774 (0.08) & 42.03$^{+0.29}_{-0.20}$ & 0.585  & 0.78 & 0.51 & $z$ from \citet{tran05}\\ 
MS2053-X5   & CXOU J205608.1-043211 & 22.73 (0.05) & 0.421 (0.08) & 42.59$^{+0.20}_{-0.15}$ & 0.588  & 2.59 & 1.71 & \\ 
\enddata
\tablecomments{ AGN in MS2053-04 ($z = 0.583$). The columns include:
(1) our identifier; (2) full name; (3) $I-$band magnitude; (4) $V-I$
color; (5) log of the hard X-ray luminosity; (6) redshift; (7)
projected distance from the cluster center in Mpc; (8) projected
distance relative to $r_{200}$; (9) notes about the X-ray source.  }
\end{deluxetable*}

The four slit masks were observed in August 2004 for three hours each
and these data were processed into 2-D spectra with the COSMOS
package\footnote{http://www.ociw.edu/Code/cosmos}. 1-D spectra were
extracted with the IRAF {\sc APALL} package and redshifts were
measured with a modified version of the Sloan Digital Sky Survey
pipeline, although each was also inspected by eye. This identified a
total of 41 counterparts to X-ray sources, including four at the
cluster redshift. To supplement these observations and obtain improved
flux measurements we reprocessed the original \chandra\ data and added
the second ACIS-I pointing. These data were processed with {\sc ciao
3.3.0.1}\footnote{\url{http://asc.harvard.edu/ciao/}}, {\sc caldb
3.2.4} and NASA's {\sc ftools 6.1.1}\footnote{
\url{http://heasarc.gsfc.nasa.gov/docs/software/lheasoft/}}. Individual
X-ray sources were identified with the {\sc ciao wavdetect} algorithm
from both the individual observations and a merged version of the
observations.  This analysis reidentifed the four X-ray counterparts
at the cluster redshift. We also cross-correlated these X-ray sources
with a spectroscopic study by \citet{tran05} and identified one
additional X-ray source at the cluster redshift. The properties of
these five sources are provided in Table~\ref{tbl:ms2053}.

\section{Results}

We have identified a total of five X-ray counterparts at the redshift
of MS2053-04, although we only classify the one AGN within $r_{200}$
as a member. The AGN classification is based on the rest-frame, hard
X-ray luminosity [2-10 keV] of $L_{X,H} = 10^{42}$ \ergs\ (measured on
the merged datasets). Of the remaining four X-ray counterparts, one is
associated with the Brightest Cluster Galaxy (BCG). This source is
coincident with the peak of the X-ray emission from the intracluster
medium and evidence for an additional, AGN component is inconclusive
\citep[see also][]{tran05}. The three remaining X-ray counterparts lie
outside the cluster's projected $r_{200}$ radius of 1.5 Mpc, which we
calculated from the cluster's velocity dispersion
\citep{carlberg97,treu03} of $\sigma = 865$ \kms\
\citep{tran05}. While they may be bound to the cluster, we adopt the
projected radius of $r_{200}$ to compare measurements of the AGN
fraction between clusters.  Figure~\ref{fig:cluster} presents the
visible-wavelength spectra, $R-$band images, and X-ray images of all
five of the X-ray sources at the redshift of MS2053-04.

\subsection{The AGN Fraction in MS2053-04} 

Calculation of the AGN fraction in MS2053-04 requires measurement of
the number of cluster members without luminous X-ray
counterparts. Following the AGN fraction measurement of
\citet{martini06}, we choose to estimate the number of cluster members
above a rest-frame $M_R < -20$ mag, which corresponds to $I = 22.4$
mag for an evolved stellar population at this redshift. As our
spectroscopic catalog is not complete for all galaxies to this
magnitude limit, we instead use the efficiency of our spectroscopic
identification of cluster members, the surface density of resolved, $I
\leq 22.4$ mag objects, and the membership data collected by
\citet{tran05} to estimate the total cluster galaxy population. We
calculate this efficiency for the full sample, as a function of
magnitude, and as a function of color and estimate the cluster galaxy
population $N_{est}$ within $r_{200}$ is 66 (no binning), 74 (color
binning), or 66 (magnitude binning). From this analysis we conclude
that there are approximately 66 galaxies brighter than $M_R < -20$ mag
in the cluster.

\subsection{Literature data and the Evolution of Cluster AGN} 


%
%
\begin{deluxetable}{lcccc}
\tablecolumns{5}
\tabletypesize{\scriptsize}
\tablecaption{Known AGN in High-Redshift Clusters \label{tbl:highz} }
\tablehead{
\colhead{Object ID} &
\colhead{$z$} &
\colhead{log $L_{X,H}$} & 
\colhead{$R/R_{200}$} &
\colhead{Refs} \\ 
}
\startdata
CL0016+1619 & & & & \\
E 0015+162 		& 0.553 & 45.48 & 0.31 & a \\ 
CXOMP J001842.0+163425 	& 0.55  & 42.59 & 0.80 & b \\
 & & & & \\
CL0542-4100 & & & & \\
CXOMP J054240.8-405626 	& 0.639 & 43.59 & 0.40 & b \\
CXOMP J054255.0-405922 	& 0.644 & 43.00 & 0.12 & b \\
CXOMP J054251.4-410205 	& 0.637 & 43.27 & 0.21 & b \\ 
CXOMP J054248.2-410140 	& 0.634 & 43.16 & 0.17 & b \\
CXOMP J054259.5-410241 	& 0.638 & 43.29 & 0.33 & b \\
 & & & & \\
CL0848.6+4453 & & & & \\
CXOSEXSI J084837.5+445710 	& 0.569 & 42.9  & 0.34 & c \\
CXOSEXSI J084843.2+445806 	& 0.566 & 42.6  & 0.34 & c \\
CXOSEXSI J084846.0+445945 	& 0.567 & 43.1  & 0.62 & c \\
CXOSEXSI J084858.0+445434 	& 0.573 & 43.8  & 0.46 & c \\
CXOSEXSI J084931.3+445549 	& 0.567 & 42.9  & 1.49 & c \\
\enddata
\tablecomments{
AGN in three additional, high-redshift clusters from the literature: 
CL0016+1609 ($z = 0.5466$); CL0542-4100 ($z = 0.630$); CL0848.6+4453 ($0.57$). 
For each AGN in column (1) we list: (2) the redshift; (3) log of the hard 
X-ray luminosity; (4) projected distance from the cluster center relative 
to $r_{200}$; (5) references are: (a) \citet{margon83}; (b) \citet{silverman05a}; (c) \citet{eckart06}. 
}
\end{deluxetable}


\begin{figure}
\plotone{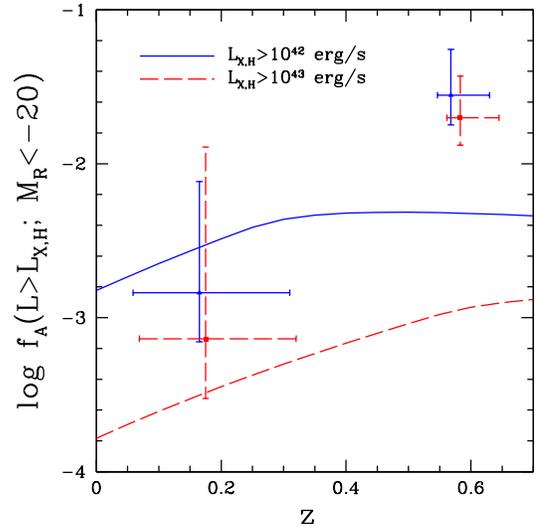}
\caption{Evolution of the cluster AGN fraction with redshift. Cluster
AGN with $L_{X,H} > 10^{42}$ \ergs ({\it blue triangles}) and $L_{X,H}
> 10^{43}$ \ergs ({\it red squares}) are more common at $\bar{z} \sim
0.6$ than in the low-redshift ($\bar{z} \sim 0.2$) sample of eight
clusters studied by \citep{martini06}. The redshift range of the
clusters used is marked by the error bars. The evolution is also more
pronounced than expected from integration of the \citet{ueda03} HXLF.
\label{fig:fraction} } 
\end{figure}

We searched the \chandra\ archive and the literature to identify
additional clusters at $z \sim 0.6$ with both comparably sensitive
observations and spectroscopic observations of the X-ray sources and
identified three clusters: CL0016+1609 ($z = 0.5466$, 61ks) and
CL0542-4100 ($z = 0.630$, 51ks), where X-ray sources in these fields
had been targeted by the ChaMP survey \citep{kim04,silverman05a} and
CL0848.6+4453 ($0.57$, 185ks), which has been targeted by the SEXSI
survey \citep{harrison03,eckart06}. In addition, CL0016+1609 has
extensive membership data and a measured velocity dispersion of
$\sigma = 1234$ \kms \citep{dressler92,carlberg96,ellingson98}. From
these membership data and the reported completeness of the CNOC
survey, we estimate that there are approximately 200 cluster members
more luminous than $M_R = -20$ mag in CL0016+1609.  The other two
clusters in this sample, CL0542-4100 and CL0848.5+4453, do not have
substantial spectroscopic data outside of the X-ray counterparts. For
these clusters, we use the measured X-ray temperatures of 7.9 keV
\citep{xue00} and 3.6 keV \citep{holden01}, respectively, to estimate
velocity dispersions of $\sigma = 1200$ \kms and $670$ \kms. We then
use the relationship between velocity dispersion and cluster richness
$N_{gals}$ from the MAXBCG survey \citep{koester07} to estimate the
number of members brighter than $M^* + 1$, or approximately our galaxy
absolute magnitude threshold. From their fitting formula we estimate
that there are 152 members in CL0542-4100 and 24 in
CL0848.5+4453. This relation also predicts 55 members in MS2053-04 and
172 members in CL0016+1609 and these values are agree with our
previous estimates to within 20\%.  We adopt these membership values
for all four clusters to consistently calculate the AGN fraction; we
discuss the uncertainties in this assumption below.

Our observations of MS2053-04, along with those for CL0016+1619 and
CL0848.6+4453, are approximately deep enough to detect sources to a
limiting rest-frame hard X-ray luminosity of $L_{X,H} > 10^{42}$
\ergs. In these three clusters we have identified a total of seven AGN
within $r_{200}$ in a total population of 251 galaxies.  For
comparison, we identified only two AGN above this limit in a sample of
eight low-redshift ($\bar{z} \sim 0.2$) clusters with a total of 1377
cluster galaxies \citep{martini07}. At $\bar{z} \sim 0.6$, we
therefore find the AGN fraction is $f_A(L_{X,H}>10^{42};M_R<-20) =
0.028^{+0.019}_{-0.012}$ (7 AGN divided by 251 cluster galaxies). The
quoted uncertainties correspond to 90\%, one-sided Poisson confidence
limits.  The data for the highest-redshift cluster in our sample
(CL0542-4100) is only sensitive to AGN more luminous than $L_{X,H} >
10^{43}$ \ergs\ and there are a total of eight AGN within $r_{200}$
above this limit in these four clusters (with an estimated $\sim 400$
total cluster members), in striking contrast to the presence of only
one comparably luminous AGN in eight low-redshift clusters. For this
X-ray luminosity threshold we estimate that the AGN fraction is
$f_A(L_{X,H}>10^{43};M_R<-20) = 0.020^{+0.012}_{-0.008}$.  In contrast, at
$\bar{z} = 0.2$ the AGN fractions from \citet{martini07} are
$f_A(L_{X,H}>10^{42};M_R<-20) = 0.0015_{-0.0011}^{+0.0024}$ and
$f_A(L_{X,H}>10^{43};M_R<-20) = 0.0007_{-0.0007}^{+0.0021}$, or
approximately a factor of 20 lower.  Figure~\ref{fig:fraction} plots
the cluster AGN fraction for these two luminosity thresholds at low
and high redshift and illustrates the significant increase in the
cluster AGN fraction at high-redshift compared to the lower-redshift
sample. For comparison, we also show the relative evolution of the
integrated space density of AGN above these two luminosity thresholds
from the parameterization of \citet{ueda03}. While the overall
normalization of the curves is arbitrary, the relative evolution of
the $L_{X,H} > 10^{42}$ \ergs and $L_{X,H} > 10^{43}$ \ergs
samples are not. These curves indicate the cluster AGN fraction
increases more rapidly with lookback time than the AGN space density
in the field. Specifically, the field AGN space density increases by
only a factor of 1.5 from $z \sim 0.2$ to $z \sim 0.6$ for $L_{X,H}
> 10^{42}$ \ergs\ and only a factor of 3.3 for $L_{X,H} >
10^{43}$ \ergs, compared to the factor of $\sim 20$ we observe in
clusters. While there is substantial variation between the AGN
fraction of individual clusters, if we exclude CL0848, the largest
contributer to the AGN fraction, and redo the calculation, we still
see a factor of 10 increase in the AGN space density.

While the substantial evolution in the AGN fraction in clusters is due
to small numbers of AGN, it is measured from a large population of
inactive cluster galaxies. The measured fractions at low and high
redshift are formally inconsistent with the 90\%, one-sided confidence
intervals. At high redshift we have more AGN and consequently have a
more significant measure of the AGN fraction. We therefore use these
measurements of the high-redshift AGN fraction and a binomial
probability distribution to ask the probability of detecting 2 AGN
with $L_{X,H} > 10^{42}$ \ergs\ or 1 with $L_{X,H} > 10^{43}$ \ergs\
in the low-redshift sample and calculate that the probability (for
each) is less than 1\%. The main systematic error in these
calculations is in the number of galaxies brighter than $M_R < -20$
mag in each cluster. The detailed membership information for MS2053-04
\citep{tran05} and CL0016+1619 \citep{dressler92,ellingson98} suggest
there are 66 and 200 members in these clusters, respectively, and both
of these values agree to within 20\% with the richness -- velocity
dispersion relationship from \citet{koester07}. This suggests that
systematic errors in membership are insignificant compared to the
observed evolution in the AGN fraction. Two additional sources of
uncertainty in these estimates are galaxy luminosity evolution between
$z \sim 0.6$ and $z \sim 0.2$ and the actual sensitivity and
completeness of the \chandra\ datasets. First, $M_R^*$ is
approximately 0.4 mag brighter at $z \sim 0.6$ than $z \sim 0.2$. Our
fixed luminosity threshold extends further below $M_R^*$ at high
redshift and therefore we have overestimated the cluster membership
(relative to $M^*$) at higher redshift and underestimated the AGN
fraction.  Secondly, the \chandra\ observations are not be uniformly
sensitive to our adopted thresholds of $L_{X,H} > 10^{42}$ \ergs\ and
$L_{X,H} > 10^{43}$ \ergs, such as the fact that the 61ks observation
of CL0016+1609 is shallower than the 90ks of total integration time
available for the center of MS2053-04.  We estimate the size of this
effect by calculating the approximate depth of each \chandra\ exposure
and integrating the hard X-ray luminosity function of \citet{ueda03}
from the current depth to $L_{X,H} > 10^{42}$ \ergs.  This exercise
indicates that we may expect as many as 50\% more AGN if these data
were uniformly sensitive to $L_{X,H} > 10^{42}$ \ergs\ sources. Two
final points are that the spectroscopy of all X-ray sources may not be
complete and it is difficult to unambiguously identify an AGN in the
BCG due to the substantial X-ray emission from the ICM. All of these
potential systematic effects would increase the AGN fraction and
strengthen our result.

\section{Summary}

We have analyzed archival \chandra\ observations of four high-redshift
$z \sim 0.6$ clusters of galaxies to derive the first measurement of
the cluster AGN fraction at high redshift. Our spectroscopic
observation of the $z = 0.586$ cluster MS2053-04 identified one
cluster AGN with a rest-frame, hard X-ray luminosity $L_{X,H} >
10^{42}$ \ergs, while when we include spectroscopic observations from
the literature we find a total of eight AGN with $L_{X,H} >
10^{43}$ \ergs\ in four clusters and seven with $L_{X,H} > 10^{42}$
\ergs\ in the three clusters with longer \chandra\ observations. We
use our spectroscopic survey of inactive galaxies in MS2053-04,
literature data for CL0016, and simple scaling arguments to estimate
the total number of inactive galaxies in each cluster above a
rest-frame luminosity of $M_R < -20$ mag to estimate the AGN fraction
in clusters above these two hard X-ray luminosity thresholds:
$f_A(L_{X,H}>10^{42};M_R<-20) = 0.028^{+0.019}_{-0.012}$ and
$f_A(L_{X,H}>10^{43};M_R<-20) = 0.020^{+0.012}_{-0.008}$.

Although these estimates correspond to a small fraction of the total
cluster galaxy population, they represent a substantial increase over
the measured cluster AGN fraction at low-redshift. Specifically, in a
sample of eight clusters of galaxies with an average redshift of $z
\sim 0.2$ \citet{martini06} measured only one AGN with $L_{X,H} >
10^{43}$ \ergs\ and only two AGN with $L_{X,H} > 10^{42}$ \ergs. From
these measurements we find that the cluster AGN fraction has increased
by approximately a factor of 20 for these two hard X-ray luminosity
thresholds between $\bar{z} \sim 0.2$ and $\bar{z} \sim 0.6$. This
evolution corresponds to a significantly greater increase in the
cluster AGN population at high-redshift than the measured evolution of
the field hard X-ray LF by \citet{ueda03} over the same redshift range
and points to the existence of an ``AGN Butcher-Oemler Effect'' in
clusters of galaxies. The overdensity of AGN at high redshift could be
an additional source of preheating of the ICM during cluster assembly,
has important cosmological implications for X-ray studies of
high-redshift clusters, and could show how the environment influences
the AGN population.

\acknowledgements

Support for this work was provided by the National Aeronautics and
Space Administration through Chandra Award Numbers 04700793 and
05700786 issued by the Chandra X-ray Observatory Center, which is
operated by the Smithsonian Astrophysical Observatory for and on
behalf of the National Aeronautics Space Administration under contract
NAS8-03060.  We greatly appreciate the excellent staffs of the Las
Campanas Observatory and the Magellan Project Telescopes for their
assistance with these observations and the helpful comments from the
referee.  This paper includes data gathered with the 6.5 meter
Magellan Telescopes located at Las Campanas Observatory, Chile.  This
research has made use of the NASA/IPAC Extragalactic Database (NED)
which is operated by the Jet Propulsion Laboratory, California
Institute of Technology, under contract with the National Aeronautics
and Space Administration.

{\it Facility:} {\facility{du Pont (Tek No.\ 5 imaging CCD, WFCCD)},
\facility{Magellan:Baade (IMACS imaging spectrograph)}


%
%

\end{document}